\documentclass[reprint, amsmath,amssymb,prfluids,prl]{revtex4-2}

\usepackage{graphicx}
\usepackage{dcolumn}
\usepackage{bm}

\newcommand{\vort}{\bm{\omega}}
\newcommand{\veloc}{\bm{u}}
\newcommand{\veloce}{\bm{u}_\eta}
\newcommand{\grad}{\bm{\nabla}}
\newcommand{\gams}{\bm{\gamma}}
\newcommand{\pos}{\bm{x}}
\usepackage{xcolor}
\begin{document}

\preprint{APS/123-QED}

\title{A toy model of turbulent shear flow using vortons }
\author{Wandrille Ruffenach}
\affiliation{ENS de Lyon, CNRS, LPENSL, UMR5672, 69342, Lyon cedex 07, France}
\affiliation{Universit\'e Paris-Saclay, CEA, CNRS, SPEC, 91191, Gif-sur-Yvette, France.}
\author{Lucas Fery}%
\affiliation{Laboratoire des Sciences du Climat et de l’Environnement, CEA Saclay l’Orme des Merisiers, UMR 8212 CEA-CNRS-UVSQ, Universit\'e Paris-Saclay \& IPSL, Paris, France,}
\affiliation{Universit\'e Paris-Saclay, CEA, CNRS, SPEC, 91191, Gif-sur-Yvette, France.}
\author{B\'ereng\`ere Dubrulle}
\affiliation{Universit\'e Paris-Saclay, CEA, CNRS, SPEC, 91191, Gif-sur-Yvette, France.}
\email{berengere.dubrulle@cea.fr}

\date{\today}

\begin{abstract}
    We introduce a novel toy model for shear flows, exploiting the spatial intermittency and the scale separation between large-scale flows and small-scale structures.
    The model is highly sparse, focusing exclusively on the most intense structures, which are represented by vortons---dynamically regularized quasi-singularities that experience rapid distortion from the large-scale shear. The vortons, in turn, influence the large-scale flow through the sub-grid stress tensor. Despite its simplicity, the model displays an interesting transition between two distinct regimes: (i) a laminar regime, where dissipation is entirely attributed to the large-scale flow, and the vortons dynamics is essentially diffusive, and (ii) a turbulent regime, in which most of the dissipation arises from the vortons.
    These regimes correspond to different scalings of dissipation and the Grashof number as functions of the Reynolds number, with power-law relationships that resemble those observed in classical turbulence.
    \begin{description}
        \item[Usage]
              Secondary publications and information retrieval purposes.
        \item[Structure]
              You may use the \texttt{description} environment to structure your abstract;
              use the optional argument of the \verb+\item+ command to give the category of each item.
    \end{description}
\end{abstract}

\maketitle

\section{Introduction}
Near-wall turbulence arises in many industrial or geophysical flows. With respect to other types of turbulence, it is peculiar in several aspects: (i) it is anisotropic; (ii) it is very intermittent, both in space and time, and both at the turbulent transition or later; (iii) it is non-local, and piloted by interactions between the large-scale shear and the small-scale turbulent structures. These three aspects make its modeling challenging,
as traditional large eddy simulations have difficulties to resolve the near-wall structures, and traditional RANS model do not capture the spatial or temporal  intermittency.

These observations motivated the introduction of non-local models, based on the two way coupling between large-scale shear and small-scale vorticity wave packets \cite{Nazarenko_2000a,Nazarenko_2000}. On the one hand, the action of large scales onto small scales is described by rapid distorsion theory (RDT), where small scales are advected and sheared by the mean shear. On the other hand, the small scales act on the large scales via the Reynolds stress. The small scales correspond to debris from coherent vortices generated in the viscous sublayer, that penetrate in the overlap region and provide a continuous forcing allowing to reach a statistically steady state.
In such setting, analytical computations are possible both in 2D and 3D, and lead to the celebrated log-law of the wall~\cite{Nazarenko_2000a,Nazarenko_2000} or to the complete description of equilibrium velocity profiles in plane parallel flows~\cite{Dubrulle_2001}. These computations are limited to the regime where turbulence is weak, so that non-linear interactions between the small-scale vorticity packets is neglected. This somehow limits the interest of the model to understand the physics of shear flows.

Including non-linear interactions is however challenging, as it may require the integration of the partial differential Navier-Stokes equations in 3D, making the corresponding model too computationally involved to be of any practical use. In this paper, we show that it is possible to avoid the appeal to partial differential equations by using Novikov's
vorton approximation, in which the small scale vorticity is described by point-like singularities, named vortons, that interact non-linearly following Biot-Savart law and the discretized inviscid Euler equation. In such setting, the partial differential equations can be replaced by a set of coupled ordinary differential equations that describe the interactions between the $N$ vortons. This provides the third mechanism that is missing in the RDT theory of~\cite{Nazarenko_2000a,Nazarenko_2000,Dubrulle_2001}, opening the way to interesting applications.
Specifically, we show in the present paper, that using only 27 vortons, we are able to reproduce several features of near-wall turbulence, such as the laminar to turbulent transition, or
the log-normal statistics of the energy dissipations. The resulting toy model, that is both sparse and computationally cheap, may then be seen as a promising tool to explore
non-linear interactions in shear flows.

\section{Novikov model}
\subsection{Equations}
Our model builds from Novikov's model, which belongs to a more general class of vortex particle methods~\cite{Cottet_2000a,Mimeau_2021a}.  In those methods, the vorticity field is discretized into localized vortices of given circulation that are advected by the flow. In two dimensions of space, vorticity is a Lagrangian invariant of Euler equation and discretizing the vorticity field naturally leads to the famous Onsager vortex gas model, predicting the condensation of energy at large scales \cite{Onsager_1949}. In three dimensions however, the vorticity stretching term in the Euler equation \eqref{eq:Euler} changes the picture.
\begin{eqnarray}
    \dfrac{\mathrm{D}\, \vort}{\mathrm{D}\, t} &=& (\vort \cdot \grad ) \veloc,\label{eq:Euler} \\
    &=& (\vort \cdot \grad ^\mathrm{T}) \veloc. \label{eq:Euler_transposed}
\end{eqnarray}
In order to find approximate solutions of Euler equations \eqref{eq:Euler} and following the ideas of the two dimensional point vortices, Novikov \cite{Novikov_1983} introduced the \textit{vortons} model. These Lagrangian vortex particles, located at position $\pos_\alpha$ and with circulation (vorticity times a volume) $\gams_\alpha$ generate a vorticity field of the form
\begin{equation} \label{eq:dirac_vorticity}
    \vort(\pos,t)= \sum_\beta \gams_{\beta} (t)  \delta  (\pos - \pos_{\beta} (t)).
\end{equation}
The velocity field is recovered using Biot-Savart law,
\begin{equation}\label{eq:velocity_biotsavart}
    \veloc(\pos,t)=-\dfrac{1}{4\pi}  {\sum_\beta} \dfrac{ \pos -\pos_ {\beta}}{\|\pos -\pos_ {\beta}\|^3} \times \gamma_ {\beta},
\end{equation}
and is by construction divergence-free.
The fields \eqref{eq:dirac_vorticity} and \eqref{eq:velocity_biotsavart} associated with Euler equation \eqref{eq:Euler} in a Lagrangian framework yield the time evolution for the variables $\pos_\alpha$ and $\gams_\alpha$,

\begin{equation}
    \left\{ \begin{aligned}
        \Dot{\pos}_\alpha(t)  & = \veloc(\pos_\alpha,t),                                  \\
        \Dot{\gams}_\alpha(t) & = \left[\grad \veloc(\pos_\alpha,t)\right]  \gams_\alpha,
    \end{aligned} \right.
    \label{eq:classical_scheme}
\end{equation}

where the velocity field $\veloc(\pos_\alpha,t)$ and its gradient $\grad\veloc(\pos_\alpha,t)$ are evaluated with \eqref{eq:velocity_biotsavart} but without the $\alpha=\beta$ term in the sum in order to avoid the unphysical behavior caused by the singularity of the model at the origin.

\subsection{Drawbacks of Novikov model and its modifications}
Novikov model presents several flaws:
\begin{itemize}
    \item The vorticity field \eqref{eq:dirac_vorticity} is not divergence free in general  and if so, will not remain under time evolution. This issue can be tackled by considering other discretizations of the vorticity field such as~\cite{mumford2012euler}.  
    \item The vorton model cannot generate a statistically steady state. Indeed, for a system of two vortons obeying \eqref{eq:classical_scheme}, Novikov \cite{Novikov_1983} showed that $\| \gams_\alpha(t) \|$ diverges exponentially with time for a given set of initial conditions.
    \item The fields generated by $\pos_\alpha$, $ \gams_\alpha$ with dynamical system \eqref{eq:classical_scheme} are not a solution (in the weak sense) of Euler equations~\eqref{eq:Euler}~\cite{Saffman_1986}.
\end{itemize}
This last problem precludes the use of Novikov model to simulate Euler equations. To address this drawback, a modification of the model was proposed, based on a remark by \cite{Winckelmans_1993}. Due to the  non-solenoidal character of the vorticity, solutions of the Novikov model are such that $\vort \neq \grad \times u$. As a consequence the Euler equation \eqref{eq:Euler} and its transposed version \eqref{eq:Euler_transposed} are not equivalent. It turns out that if we now keep $\vort$ and $\veloc$ given by \eqref{eq:dirac_vorticity} and \eqref{eq:velocity_biotsavart}, using the transposed scheme, the system
\begin{equation}
    \left\{ \begin{aligned}
        \Dot{\pos}_\alpha(t)  & = \veloc(\pos_\alpha,t),                                               \\
        \Dot{\gams}_\alpha(t) & = \left[\grad \veloc(\pos_\alpha,t)\right] ^{\mathrm{T}} \gams_\alpha,
    \end{aligned} \right.
    \label{eq:transposed_scheme}
\end{equation}
yields a weak solution \cite{Winckelmans_1988} of the Euler equations \eqref{eq:Euler_transposed} and conserves key physical quantities such as the total vorticity \cite{Winckelmans_1993} or helicity. \

However, this modification itself still presents two drawbacks:
\begin{itemize}
    \item the transposed vorton model cannot generate a statistically steady state. Indeed, like in the original vorton model with $N\geq 2$ vortons a random initial condition, the quantity
          \begin{equation} \label{eq:L2_norm_vort}
              \Gamma(t)= \sum_{\alpha= 1}^N \| \gams_\alpha(t)\|^2
          \end{equation}
          would most surely be divergent in time, making any turbulent like stationary regime unreachable.
    \item In its original philosophy, this model is just an alternative way to solve the Euler equations. The equivalence between this method and the Euler equation is then guaranteed only in the limit where one can add indefinitely many vortons in the system as time passes
          by and vortex filaments are getting stretched, resulting in vortons moving apart far from each other, breaking the continuous line approximation. \
\end{itemize}

These difficulties combined with the flaws mentioned earlier restricted the use of this kind of models to simple situations such as vortex lines interactions or vortex rings leap-frogging for instance~\cite{Novikov_1983}.

\section{A new model for shear flows}
\label{sec:model}
\subsection{General picture}
The new model we consider is based on several ideas, that are meant to correct the main drawbacks identified earlier.

\begin{itemize}
    \item {\sl Idea} $\#$1: {\bf Sparsity}. Instead of considering vortons as elementary blocks used to decompose the whole vorticity field---which implies the consideration of many vortons---we instead consider that vortons model the few extreme events of vorticity arising in the flow, extremes that will be the main players to pilot the full flow dynamics. In that approximation, we can keep only a few vortons as time passes by, as extreme events are by definition isolated and rare events.

    \item  {\sl Idea} $\#$2: {\bf Regularization}. To avoid runaway of the vorton energy and allow for stationary states, we introduce an effective size for the vortex cores, this is done by mollifying the vorticity field
          \begin{eqnarray}
              \omega_\eta(\pos,t) &=& \left(\zeta_\eta \ast \omega \right)(\pos,t) \nonumber\\
              &=& {\sum_\beta} \zeta_\eta(\pos-\pos_{\beta}(t)) \gams_{\beta}(t),
          \end{eqnarray}
          where
          \begin{equation}
              \zeta_\eta(\pos)= \dfrac{1}{\eta^3} \zeta\left( \dfrac{\pos}{\eta} \right),
          \end{equation}
          is an approximation of Dirac mass at scale $\eta$. Several choices can be made for $\zeta$, see for instance~\cite{Winckelmans_1993}and~\cite{choquin1988analyse} for the proof of convergence of such regularized models . In the following, we will use the so-called low order algebraic kernel,
          \begin{equation}
              \zeta(\pos)=\dfrac{3}{4\pi |\pos|_1^5}
          \end{equation}
          with the pseudo-norm $ |\pos|_L=\sqrt{\|\pos\|^2+L^2}$. Using the low order algebraic kernel, the regularized vorticity $\vort_\eta$ and velocity $\veloc_\eta$ now writes
          \begin{eqnarray}
              \vort_\eta(\pos,t)  &=& \sum_{\alpha=1}^{N} \gams_\alpha \dfrac{3\eta^2}{4\pi |\pos|^5_\eta},                                          \\
              \veloc_\eta(\pos,t) &=& -\dfrac{1}{4\pi} \sum_{\alpha=1}^{N}  \dfrac{\pos-\pos_\alpha}{|\pos-\pos_\alpha|^3_\eta}  \times\gams_\alpha.
          \end{eqnarray}

           With this choice for $\zeta$, the shape of the velocity field given by \eqref{eq:velocity_biotsavart} remains the same appart from the norm Euclidean norm $\| \cdot\|$ which changes to the pseudo-norm $| \cdot |_\eta$.

    \item  {\sl Idea} $\#$3: {\bf Self-interactions}. We assume that the vortons interact with each other following the transposed version of Euler equations. Using the regularization, the transposed scheme written in terms of $\bm{r}_{\alpha \beta}= \pos_\alpha - \pos_\beta$ and $\gams_\alpha$ is
          \begin{equation}
              \begin{aligned}
                  \Dot{\pos}_\alpha  & = -\dfrac{1}{4\pi} \sum_{\beta=1}^N \dfrac{\bm{r}_{\alpha \beta}}{|\bm{r}_{\alpha \beta}|_{\eta}^3} \times \gams_\beta                                           \\
                  \Dot{\gams}_\alpha & = -\dfrac{1}{4\pi} \sum_{\beta=1}^N \biggl( \dfrac{\gams_\alpha \times \gams_\beta}{| \bm{r}_{\alpha \beta} |^3_\eta}                                            \\
                                     & -\dfrac{ 3\bm{r}_{\alpha \beta}}{|\bm{r}_{\alpha \beta}|^5_\eta} \left( \bm{r}_{\alpha \beta}\cdot \left( \gams_\alpha \times\gams_\beta \right)\right) \biggr).
              \end{aligned}
          \end{equation}
          The new parameter $\eta$ corresponds to the size of vortices. Allowing this new, unconstrained variable to be time dependent makes it possible to balance the vortex stretching term, hence reaching a statistically stationary regime.

    \item  {\sl Idea} $\#$4: {\bf Dynamical regularization}. We assume that the regularization length scale is dynamically ajusted, and can either increase under the effect of viscous diffusion or decrease because of vortex stretching. The impact of such dynamics can be obtained by integrating the Navier-Stokes equations written for vorticity,
          \begin{equation}
              \dfrac{\mathrm{D} \, \vort  }{\mathrm{D} \, t}= \left( \vort \cdot \grad^{\mathrm{T}} \right) \veloc+ \nu \Delta \vort,
          \end{equation}
          over a ball centered on $\pos_\alpha$ with a radius $\epsilon \ll \eta$ and allowing for a time dependent $\eta$. Keeping only the dominant terms in the sums and considering $\Delta \psi_\eta (\pos) \approx \delta(x)$ we end up with
          \begin{eqnarray}
              \Dot{\pos}_\alpha  &=& \veloc_\eta(\pos_\alpha,t), \\
              \Dot{\gams}_\alpha &=& 3\left( \dfrac{\Dot{\eta}}{\eta}-5 \dfrac{\nu}{\eta^2} \right) \gams_\alpha + \left[ \grad\veloc_\eta(\pos_\alpha,t) \right]^{\mathrm{T}} \gams_\alpha .\label{eq:DotGamma}
          \end{eqnarray}
          Viscosity therefore has a damping effect and introducing a time dependence on $\eta$ is a way of introducing a new free parameter and therefore a supplemental constraint on the dynamics. The time evolution obtained for the vortex strength~\eqref{eq:DotGamma} is very similar to what was obtained in~\cite{alvarez2024stable} in the inviscid case.

    \item  {\sl Idea} $\#$5: {\bf Rapid distorsion by a large-scale flow}. We consider that the vortons are embedded within a large-scale flow $\bm{U}$, with velocity strain tensor $\grad \bm{U}$. Then, we take into account the action of the large-scale flow on the vortons through an additional advection by the large-scale velocity field and additional stretching by the large-scale strain-rate tensor. In final, the vortons are then  advected by the field $\veloc = \veloce + \bm{U}$ and stretched by $\left[\grad \veloc\right]^{\mathrm{T}} =  \left[\grad \veloce\right]^{\mathrm{T}} + \left[\grad \bm{U}\right]^{\mathrm{T}} $.

    \item  {\sl Idea} $\#$6: {\bf Feedback on the large-scale flow}. We consider the feedback of the vortons on the large-scale flow via the subgrid stress tensor $\tau_\ell$, where $\ell$ is a yet unspecified filtering scale.
\end{itemize}

\subsection{Application to shear flow}
\subsubsection{Equations}
\label{sec:shear_flow}
The subgrid model described above is very general. We now provide an application of the method to the transition to turbulence in shear flow.
We consider the simplest possible shear flow, given by the velocity field:
\begin{equation}
    \bm{U}(\pos,t) =a(t) \begin{bmatrix}
        \sin \left( k_s z\right) \\
        0                        \\
        0
    \end{bmatrix},
\end{equation}
where $k_s = \dfrac{2\pi}{L_s}$.
$\bm{U}(\pos,t) $ is divergence-free and is compatible with periodic boundary conditions used in numerical simulations. For simplicity, we assume that the time dependence of the large scale flow is fully encoded in its amplitude $a(t)$ while its spatial shape remains the same.We will provide a dynamics for the amplitude $a$ which involves viscous dissipation, energy exchange with the small scale vortons and a smooth random forcing term, needed to reach a statistically steady state. We will take a forcing of the form $\bm{f}(\bm{x},t) = (2/L^3) f(t) \mathrm{sin}(k_s z) \mathbf{e}_x$ where $f(t)$ is a smooth random Gaussian process of average $0$ and standard deviation $f_0$.

We can then write the equation governing the dynamics of vorticity filtered at a scale $ \ell$. We do so by convoluting the curl of Navier-Stokes equations with a mollifier $G_\ell(\pos)= 1/\ell^3G(\pos/\ell)$ to obtain :
\begin{equation}\label{eq:vorti_filtered}
    D_t \vort_\ell= \mathrm{S}_\ell \vort_\ell + \grad \times \left[\grad \cdot \tau_\ell\right]+ \nu \Delta \vort_\ell + \grad \times \bm{f}_\ell,
\end{equation}
where $\mathrm{S}_\ell$ is the filtered shear stress, $\tau_\ell = \veloc_\ell \otimes \veloc_\ell - \left(\veloc \otimes \veloc\right)_\ell$ is the sub-grid stress tensor. Exploiting the scale separation between the vortons scale $\eta$ and the shear scale $L_s$, we can choose $\eta \ll \ell \ll L_s$, allowing for the approximations,
\begin{eqnarray*}
    \bm{f}_\ell &\approx&  \bm{f}, \\
    \bm{U}_\ell &\approx&  \bm{U}, \\
    \vort_\ell  &\approx& \grad \times \bm{U}.
\end{eqnarray*}
Under the previous assumptions, the 3D filtered Navier-Stokes equations \eqref{eq:vorti_filtered} satisfied by the large-scale flow reduce to a scalar equation for its amplitude $a(t)$.
To obtain the time evolution of $a(t)$, we project equation \eqref{eq:vorti_filtered} onto the large-scale vorticity $\vort_\ell = a(t) k_s \mathrm{cos}\left(k_s z\right) \bm{e}_y$, integrating over a square box of volume $ |\mathrm{V}|=L^3$ with $L=nL_s$, $n\in\mathbb{N}^*$ and rescaling by the factor $k_s^{-1} (L^3/2)^{-1}$.
This yields the following equation for $a(t)$:
\begin{eqnarray}
    \Dot{a} &=&- \dfrac{a}{\tau_\nu^s}  -  \dfrac{3\pi^2 \theta}{32 L^3 L_s \eta} \sum_{\alpha=1}^N \gamma_{\alpha,x} \gamma_{\alpha,z} \cos\left(k_s z_\alpha\right) \nonumber\\
    && +\dfrac{2}{L^3}f(t),
    \label{aequation}
\end{eqnarray}
where $\theta \in [0,1]$ is a coupling parameter  depending on the choice of the filtering function used in \eqref{eq:vorti_filtered} and $\tau_\nu^s= (\nu k_s^2)^{-1}$ is the large-scale viscous time. The derivation of the second term on the right-hand side, representing the contribution of the subgrid stress tensor, is given in Appendix \ref{app:subgrid_stress}. Assuming the shape of $\bm{U}$ remains the same through time evolution is a very strong hypothesis. It is however possible to generalise the dynamics by considering the incompressible Galerkin truncated field
$$  \bm{U}(\pos,t) = \sum_{|\bm{n}|<M} \bm{a}_{\bm{n}}(t) e^{ i k_s \bm{n} \cdot \pos }.  $$
The time evolution of the fourier modes $\bm{a}_{\bm{n}}(t)$ will be derived in the same fashion as we did for $a$. <the cutoff $M$ must be chosen such that the large scale separation hypothesis is still valid or in other words $ Mk_S \eta \ll 1 $. In this article, we keep the toy model as simple as possible and therefore keep this idea for future work. We mention in addition that one could consider the large scale field $\bm{U}$ to be the solution of a Large Eddy Simulation. The embedding of vortex particles in LES flow has been studied in \cite{kornev2019large}.

Taking into account  the rapid distorsion of vortons by the large-scale shear flow, we then obtain the system of equations governing the vorton dynamics:
\begin{eqnarray}
    \Dot{\pos}_\alpha &=& \veloc(\pos_\alpha,t) + \bm{U}(\pos_\alpha,t),
    \label{eq:x_dot} \\
    \Dot{\gams}_\alpha &=& 3\left( \dfrac{\Dot{\eta}}{\eta}- 5\dfrac{\nu}{\eta^2} \right) \gams_\alpha + \left[\grad \veloc(\pos_\alpha,t)\right]^{\mathrm{T}} \gams_\alpha \nonumber \\
    && + \left[\grad \bm{U}(\pos_\alpha,t)\right]^{\mathrm{T}}  \gams_\alpha.
    \label{eq:g_dot}
\end{eqnarray}
In order to close this system, we have to provide the time evolution for $\eta$.

\subsubsection{Closure for the regularization length}
The closure of the system of equation is based on a kinetic energy budget between vortons and the large-scale flow. The (approximate) vorton kinetic energy is computed in Appendix \ref{app:energy}. It is given by:
\begin{equation}
    K_v= \dfrac{\Gamma}{64\eta},
    \label{VortonKE}
\end{equation}
where $\Gamma$ is defined in \eqref{eq:L2_norm_vort}.
On the other hand, the kinetic energy of the shear flow integrated over the box of volume $|\mathrm{V}|=L^3$ is,
\begin{equation}\label{DefiKs}
    K_s= \dfrac{a^2 L^3}{4}.
\end{equation}
In the inviscid and non-forced case, kinetic energy should be exchanged between vortons and shear but conserved overall.
We consider then two limiting situations: in the very viscous limit $\nu\gg 1$, the regularization  length is just set by viscosity, so that on dimensional ground $\Dot{\eta}\sim \nu/\eta$, like the core-spreading method~\cite{Cottet_2000a} (Sec. 5.6.2). In the inviscid limit $\nu\to 0$, the regularization length scale is  evolving in order to keep $K=K_s+K_v$ constant. Patching the two behaviours, we get the following equation for $\eta$:
\begin{eqnarray}
    \Dot{\eta}&=& 2 \delta \dfrac{\nu}{\eta} - \dfrac{2\eta}{5\Gamma}\biggl[ \left \langle \gams |\grad \veloc ^{\mathrm{T}} \gams \right \rangle \nonumber \\
    && +\left(1-\dfrac{3 \pi}{4}\theta \right)\left \langle \gams | \grad\bm{U}^{\mathrm{T}} \gams \right \rangle\biggr],
    \label{eq:eta_dot}
\end{eqnarray}
where $\delta$ is a free parameter and we used the shorthand notation:
\begin{equation}
    \left \langle \gams | A \gams \right \rangle = \sum_{\alpha=1}^N \gams_\alpha^{\mathrm{T}}A(\pos_\alpha,t) \gams_\alpha,
\end{equation}
for any tensor field $A(\pos,t)$.

\subsubsection{Choice of the parameters}
There are two free parameters in the model, $\theta$ and $\delta$.
\begin{itemize}
    \item The parameter $\theta\in [0,1]$ is a coupling parameter. We have no physical argument to select a particular value, so we use a choice that simplifies the equations. Namely, we will take $\theta= 4/3\pi$ to cancel the contribution from the large-scale field to the dynamics of the regularization length \eqref{eq:eta_dot}. This choice simplifies the model and makes the interaction between the large-scale field and the vortons independent of the choice of the scale filter.

    \item The parameter $\delta$ controls the viscous decay of the regularization length.  In the case $\delta=5/2$, viscous diffusion is entirely accounted for by the spreading of the vortex core, while the vortons intensities $\gamma_\alpha$ are not affected \eqref{eq:g_dot}. In the other limiting case, $\delta=0$, the vortex core size is not affected by viscosity, viscous dissipation thus only affects the vortons intensities. Thus, we should have $\delta \in [0, 5/2]$.
\end{itemize}

\subsubsection{Control parameters}
As we will see in the following sections, the statistics of the model are governed by the two physical input parameters: the viscosity $\nu$ and forcing amplitude $f_0$. These parameters can be used to build a dimensionless number usually referred to as the Grashof number
\begin{equation}\label{eq:def_grashof}
    \mathrm{Gr}= \dfrac{2 f_0}{\nu^2}.
\end{equation}
Then, to study the hydrodynamics of the model, we can define a Reynolds number based on the fluctuations of the large scale field,
\begin{equation}
    \mathrm{Re}=\dfrac{\sigma_a L_s}{\nu},
\end{equation}
where $\sigma_a^2= \mathbb{E} a^2$ is the variance of the shear flow amplitude, depending on $f_0$ and $\nu$. This definition only makes sense if $a$ reaches a statistically steady state. This will indeed be the case in both the laminar and turbulent states described later.\

\subsection{Diagnostics and Global quantities}
Diagnostics will be made using several global quantities based upon $a$ and  the fields $\veloc_\eta$ and $\vort_\eta$. A priori, these global quantities depend on $\eta$ and on the configuration $\left \lbrace  \pos_\alpha, \gams_\alpha\right \rbrace$. In practice, observables with a quadratic dependence on $\veloc_\eta$ and $\vort_\eta$ will depend at first order on $\Gamma= \sum_\alpha \| \gams_\alpha\|^2$ and $\eta$. This was indeed the case for $K_v$ given by Eq. \eqref{VortonKE}  This is the case for instance for:
\begin{itemize}
    \item the vortons dissipation rate that can be computed using equations \eqref{eq:x_dot}, \eqref{eq:g_dot} and \eqref{eq:eta_dot} as
          \begin{equation}
              \Dot{K}_v=  \left( 3 - \delta \right) \dfrac{5}{32} \dfrac{\Gamma}{\eta^3}.
              \label{eq:dissipationvort}
          \end{equation}

    \item the global energy dissipation $\Dot{K}$, that can be computed using equations \eqref{aequation} and \eqref{eq:dissipationvort}  as
          \begin{equation}
              \Dot{K}= f(t)a(t) - \nu \left[2 \pi^2 a^2 \dfrac{L^3}{L_s^2}   +  \left( 3 - \delta \right) \dfrac{5}{32} \dfrac{\Gamma}{\eta^3}\right].
              \label{eq:dissipation}
          \end{equation}
          We see that since  as  $\delta<5/2$ the contribution of vortons in the total energy budget \eqref{eq:dissipation} is negative.

    \item the vortons enstrophy
          \begin{equation}
              \Omega \equiv \int \bm{\omega}_\eta^2(\bm{x},t)^2 \, d^3x \approx \dfrac{45}{1024} \dfrac{\Gamma}{\eta^3}.
              \label{eq:omegaeta2}\end{equation}

    \item the vortons mean squared velocity gradient
          \begin{equation}
              S^2\equiv  \int \left( \nabla u_\eta (\bm{x},t) \right)^2 \, d^3x \approx \dfrac{15}{128} \dfrac{\Gamma}{\eta^3}. \label{eq:ueta2}
          \end{equation}
\end{itemize}
Here, it is important to emphasize that since $\bm{\omega}_\eta$ is not divergence-free, the vortons enstrophy and their mean squared velocity gradient do not coincide, unlike in solutions of the Navier-Stokes equations.

\section{Results}
\label{sec:results}
\subsection{Parameters}
In this section, we present the results of simulations of the model described previously.
We simulate $N=3^3 =27$ vortons in a periodic box of size $L=L_s=1$. The vortons are initially placed on a regular cubic lattice with $3\times 3 \times 3$ points spaced by $h=L/N^{1/3}$.
The forcing is chosen as:
\begin{eqnarray}
    f(t) &=& f_0 \Biggl(\alpha_0 + \sqrt{2} \sum_{j=1}^m \biggl[ \alpha_j \mathrm{cos} \left( 2\pi j\frac{t}{T_f} \right) \nonumber\\
        &&  + \beta_j \mathrm{sin} \left( 2 \pi j \frac{t}{T_f} \right)\biggr]\Biggr),
\end{eqnarray}
where $T_f$ is the forcing period, $m=\lfloor T_f/\lambda \rfloor$, with $1/\lambda$ the maximum frequency of the forcing. The random part of the forcing comes from $\alpha_i$, $\beta_i$, which are independent and identically distributed Gaussian variables of average 0 and standard deviation $1/(2m+1)$. In the end, the forcing term is smooth given the finite number of modes, $T_f$-periodic and of standard deviation $f_0$.\

Regarding the parameters, we adopt $\theta=4/3\pi$ and $\delta=9/4$. This last value is chosen so that the dissipation rate of the vortons \eqref{eq:dissipationvort}  is indeed given by $\nu S^2$, with $S^2$ given by  \eqref{eq:ueta2}.
Other parameters are taken as $\lambda=1$, $T_f=3000\lambda$. The intensities of the vortons are initialized taking their components as independant random variables distributed uniformly in $[-I,I]$ with $I=10$. The simulation time is taken as $T_f$. We integrate the model using the standard Runge-Kutta method of order 5(4) implemented within the SciPy Python library.

\subsection{Laminar and turbulent regimes.}

\begin{figure}
    \centering
    \includegraphics[scale=1.]{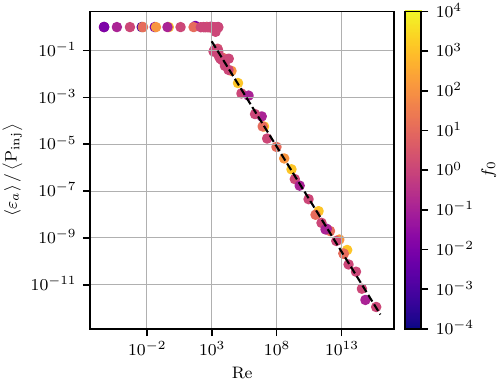}
    \caption{Time averaged dissipation rate for the large-scale flow normalized by the average injection rate as a function of the Reynolds number. The transition from a laminar state to a turbulent one occurs at $ \mathrm{Re}_c\approx 10^3$ (or $ \mathrm{Gr}_c\approx 2\times 10^6$). For $\mathrm{Re} \le\mathrm{Re}_c$ the injected power is dissipated by the large scale flow while in the turbulent regime, energy is dissipated at smaller, vortons scale. The ratio of the large-scale dissipation rate to the injected power decreases in the turbulent regime following a power law with fitted exponent $\alpha=-0.90$.}
    \label{fig:Transition}
\end{figure}

\begin{figure*}
    \centering
    \includegraphics[scale=1.]{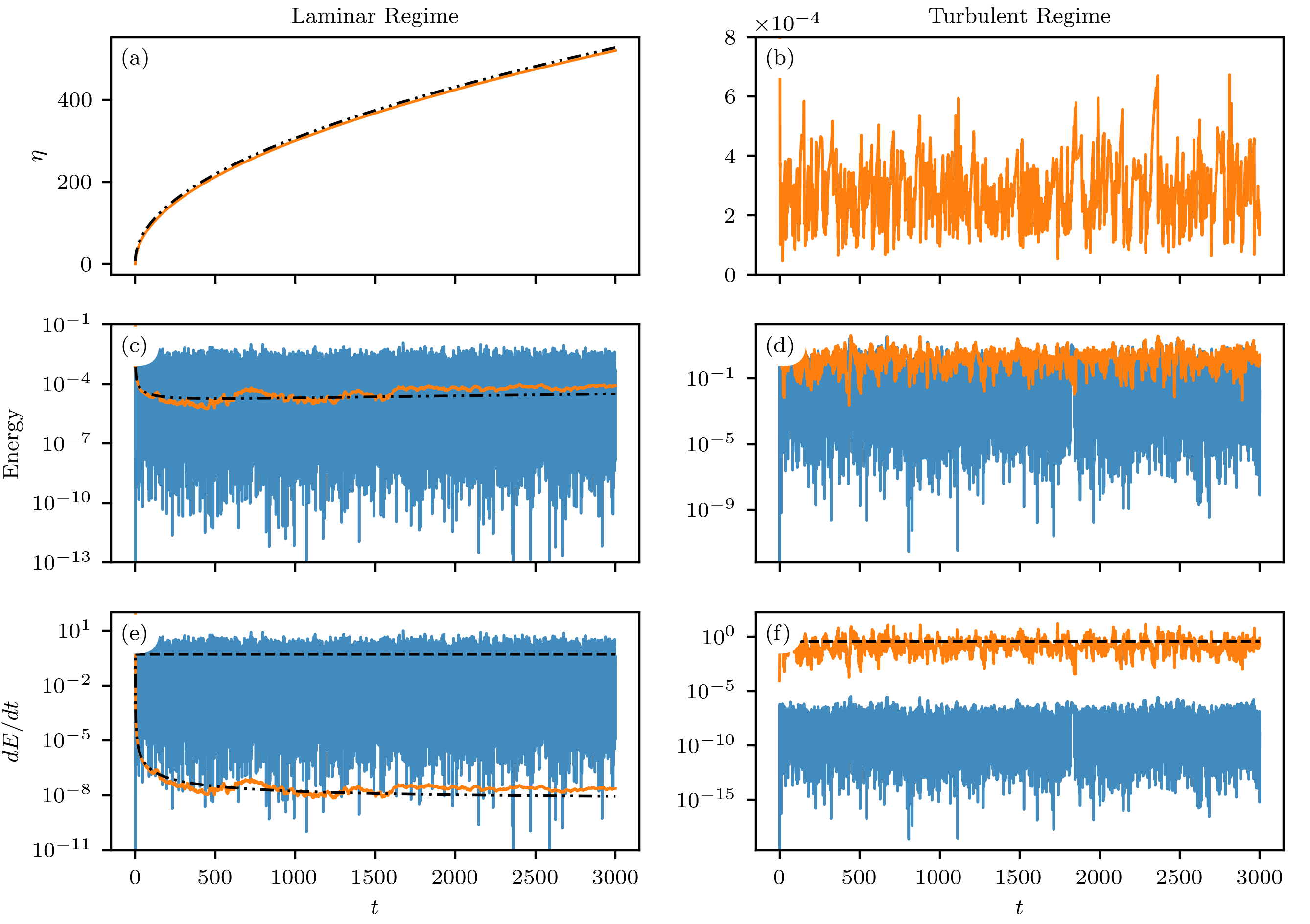}
    \caption{Time series of regularization length (a, b), kinetic energy (c, d) and viscous dissipation and injected power (e, f) in the laminar (left column: a, c, e) and turbulent (right column: b, d, f) regimes. The two regimes are both illustrated by the results of a single simulation with control parameters $\nu=10$ (respectively $\nu=10^{-9}$) and $f_0=10$ (respectively $f_0=1$) for the laminar (turbulent) regime. The first row (a,b) shows the time series of the regularization length. In the laminar regime (a), its dynamics matches the prediction $f_1(t)=\eta_0 \psi(t)$ (represented by the dash-dotted black line, slightly offset for readability) while it reaches a stationary mean value in the turbulent regime (b). The second row (c,d) shows the time series of both the large-scale kinetic energy (blue line) and of the vortons kinetic energy (orange line). In the laminar regime (c), the kinetic energy of the vortons follows well the expected dynamics at large timescales represented by the dash-dotted black line ($f_2(t) = (1/64\eta_0)(I^2/4)\psi(t)^{\frac{(5 \delta - 15)}{\delta}}(1+t(t+T_f)/((2m+1)\tau^2_\Gamma))$). The last row (e,f) depicts the energy budget with the dissipation from the large-scale flow (blue line), that from the vortons (orange line) and the time-averaged injected power (dashed horizontal black line). In the laminar regime, energy injection is mostly balanced by dissipation at large scale while in the turbulent regime, energy injection is balanced by dissipation at small scales, i.e. by the vortons. Additionally, the dissipation from vortons matches well with the expected dynamics at large timescales represented by the dash-dotted black line ($f_3(t) = (15/128\eta_0^3)(I^2/4)\psi(t)^{\frac{(3 \delta - 15)}{\delta}}(1+t(t+T_f)/((2m+1)\tau^2_\Gamma))$)}
    \label{fig:time_series}
\end{figure*}

An interesting feature of this model is the existence of a transition between two regimes, highlighted by the ratio of mean energy dissipation at large scale $\left<\varepsilon_a\right>  = 2\pi^2 \nu L^3/L_s^2 \left< a^2\right>$ to the mean injected power $\left<\mathrm{P_{inj}}\right> = \left<a(t)f(t)\right>$, as a function of the Reynolds number, which is shown in Fig. \ref{fig:Transition}. At low Reynolds number  ($Re<Re_c \approx 10^3$), this ratio is close to one, showing that all the dissipation is provided by the large-scale shear. This is confirmed by time-series shown in  Fig. \ref{fig:time_series}e). In this regime, the regularization length grows continuously as the square root of time  (Fig. \ref{fig:time_series}a). This means that the dynamics is mostly diffusive. In analogy with classical turbulence, we can identify this regime as laminar. In this laminar regime, the diffusive effects are dominating the whole dynamics allowing some approximations. The way we define the laminar regime is thus given by the set of equations \eqref{eq:LamiA} to \eqref{eq:LamiGam} investigated in the following. At the critical Reynolds number $Re=Re_c $, there is a sudden drop in the ratio, followed by a new regime where the ratio is close to $0$. This means that in this regime ($Re>Re_c$), energy is mostly dissipated by the vortons, as confirmed again in the time series shown in  Fig. \ref{fig:time_series}f). In this regime, the regularization length scale reaches a statistically stationary value (Fig. \ref{fig:time_series}b). We call this regime the turbulent regime.

\begin{figure}
    \centering
    \includegraphics[scale=1.]{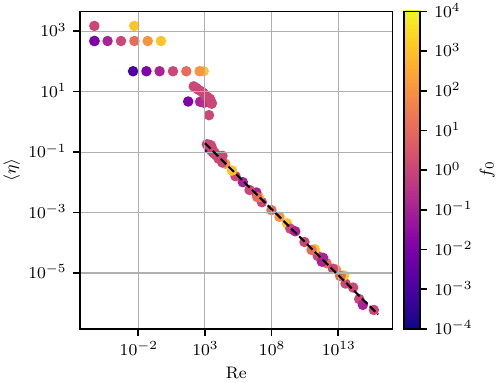}
    \caption{Mean regularization length as a function of the Reynolds number. Each dot corresponds to a single simulation, with given values of viscosity $\nu$ and forcing amplitude $f_0$ ($f_0$ being represented by the dot color). Two regimes can be identified, separated by a critical Reynold number $\mathrm{Re}_c\sim 1000$ where the regularization length either grows as the square root of time in the laminar regime (low Reynolds numbers) or reaches a stationary mean value in the turbulent regime (high Reynolds numbers). The black dashed line is a fit of a power law $<\eta>\sim \mathrm{Re}^\alpha$ in the turbulent regime, with fitted exponent $\alpha=-0.44$. For $\mathrm{Re}\leq \mathrm{Re}_c$, $\eta$ behaves as $\sqrt{\eta_0^2+4(1-\delta) \nu t}$. The finite values observed here are due to the finite time of the simulation and are indeed independent of the forcing amplitude $f_0$.}
    \label{fig:eta}
\end{figure}

\begin{figure}
    \centering
    \includegraphics[scale=1.]{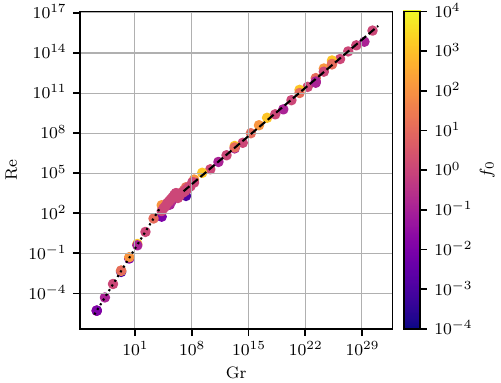}
    \caption{Reynolds number $\mathrm{Re}=\sigma_a L_s/\nu$ as a function of Grashof number $\mathrm{Gr}=2 f_0/\nu^2$. Each dot corresponds to a single simulation, with given values of viscosity $\nu$ and forcing amplitude $f_0$ ($f_0$ being represented by the dot color). The dotted and dashed lines are power law fits of the two regimes (laminar and turbulent). The fitted exponents are $\alpha=0.99$ for the laminar regime and $\alpha = 0.52$ for the turbulent regime. The scaling for the laminar regime match the prediction \eqref{eq:REvsGR_lam}.}
    \label{fig:REvsGR}
\end{figure}

The laminar to turbulent transition is also observed in the behavior of other global variables, such as the mean regularization length as a function of Reynolds number, see Figure \ref{fig:eta}. In the laminar regime, its value is mainly constrained by the finite simulation time $T_f$ as it grows continuously, while in the turbulent regime, it scales as a power law  $\left<\eta\right> \sim \mathrm{Re}^\alpha$ with $\alpha \approx -0.44$. We also observe the transition in the relation between
Reynolds number scaling and the Grashof number shown in Fig. \ref{fig:REvsGR}. At low value of Reynolds and Grashof, the scaling is linear, while
at higher values of the Grashof number, the scaling is different with $\mathrm{Re}\sim \mathrm{Gr}^\alpha$ with $\alpha\approx0.52$. In steady state turbulence, it is observed \cite{Frisch_1995} that the variance of the velocity field becomes independent
of viscosity in the limit of vanishing viscosity. In our model, this would be verified if $\alpha=1/2$ as if $\mathrm{Re}\sim \mathrm{Gr}^\alpha$ we should have $\sigma_{a} \sim f_0^\alpha \nu^{1-2\alpha}$. As shown in Fig. \ref{fig:REvsGR}, the scaling exponent fitted on our simulations results is close to $1/2$ but slightly larger (the
$95\%$ confidence interval does not include $1/2$).

\subsection{Dynamics in the laminar regime}
We can further explore the dynamics of the model in the laminar regime by simplifying the equations, neglecting the vortons-vortons interactions and the retroaction of the vortons on the large-scale flow. The system then reduces to
\begin{eqnarray}
    \Dot{a}&=&-\dfrac{ a}{\tau_\nu^s} + \dfrac{2}{L^3} f(t) \label{eq:LamiA},\\
    \Dot{\eta} &=& 2 \delta \dfrac{\nu}{ \eta},\label{eq:LamiEta}\\
    \Dot{\pos}_\alpha &=& \bm{U}(\pos_\alpha,t),\label{eq:LamiX} \\
    \Dot{\gams_\alpha} &=& 3 (2\delta - 5) \dfrac{\nu}{\eta^2}\gams_\alpha + \grad \bm{U}^{\mathrm{T}}(\pos_\alpha,t)\gams_\alpha.\label{eq:LamiGam}
\end{eqnarray}

In order to simplify future computation, we will also consider that the large scale flow is overdamped, yielding,
$$a(t)=a_{rms} \left(\frac{f(t)}{f_0} \right)$$
with $a_{rms}= (1/4\pi^2)(L_s^2/L^3) (2f_0/\nu) $ the standard deviation of $a$. Under these assumptions, the Reynolds number should thus behave in the laminar regime as,
\begin{equation}\label{eq:REvsGR_lam}
    \mathrm{Re}= \left(\dfrac{L_s}{L}\right)^3 \dfrac{\mathrm{Gr}}{4\pi^2}.
\end{equation}

The scaling derived in equation \eqref{eq:REvsGR_lam} is indeed consistent with the numerical results shown in Fig. \ref{fig:REvsGR}, with the Reynolds number scaling as the Grashof number for $\mathrm{Gr}< \mathrm{Gr_c}\approx 10^5$.

Time dependence of various physical quantities, such as the regularization scale, and the forcing terms are computed in Appendix \ref{app:laminar_regime}.  Using these estimates, we find that
the vortons kinetic energy $K_v=\Gamma /(64 \eta)$ , energy dissipation $\Dot{K}_v=\left( 3 - \delta \right) \dfrac{5}{32} \dfrac{\Gamma}{\eta^3}$ or enstrophy $\Omega_v=45 \Gamma/(2048 \eta^3)$, are thus respectively behaving as $t^{\frac{9\delta-15}{2\delta}}$,
$t^{\frac{7\delta-15}{2\delta}}$ and $t^{\frac{7\delta-15}{2\delta}}$ when $t \to \infty$. Energy would thus decrease with time in the laminar regime if $\delta < 15/9$ while energy dissipation or enstrophy would decrease if $\delta<15/7$. With our choice of $\delta=9/4$, both $K_v$ and $\Dot{K}_v$ should increase indefinitely at large times which is unrealistic. However, one should keep in mind that the regularization length $\eta$ increases indefinitely and eventually becomes larger than the size of the periodic box $L$ which is not physical. As $\eta$ increases, the vortons field is not concentrated at small scales anymore (the cutoff wave number in the energy spectrum is $1/\eta$). Consequently, the increase of the vortons kinetic energy is likely due to the progressive accumulation of energy at larger and larger spatial scales.
We represent the expected dynamics at large timescales on Fig. \ref{fig:time_series}c,d) for the energy of the vortons and their energy dissipation. These laws are indeed well verified.

\subsection{Dissipation rate}
One of the    main hypothesis of Kolmogorov turbulence phenomenology~\cite{Frisch_1995} is that the energy dissipation rate becomes independent of viscosity in the limit of vanishing viscosity, and that it should scale with the cube of the standard deviation of the velocity field. This is indeed observed in numerical simulations~\cite{Sreenivasan_1984} or experiments like the von Karman flow~\cite{Dubrulle_2019}. In the case of shear flows, this scaling depends on the boundary conditions, via the state of the surface~\cite{Eyink_2024}: for rough surfaces, this scaling is indeed observed, while for smooth surfaces, present data only evidence a slow decay with with decreasing viscosity, possibly corresponding to logarithmic corrections. \

\begin{figure}
    \centering
    \includegraphics[scale=1.]{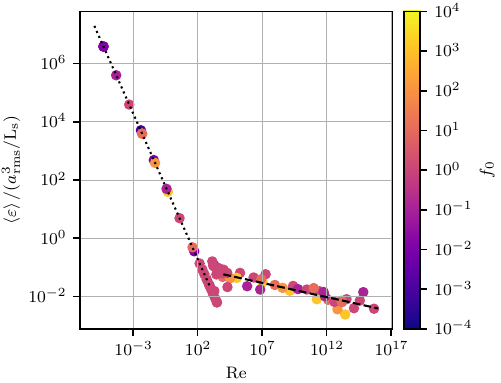}
    \caption{Rescaled total dissipation rate as a function of the Reynolds number. Similarly to turbulent flow, the dissipation rate scales as the cube of the velocity field variance. In the laminar regime, the rescaled dissipation scales as $\mathrm{Re}^{-1}$ (the fitted coefficient is $\alpha=-1.00$), while in the turbulent regime, we get a smaller exponent ($\alpha = -0.10$), the data points being quite scattered.}
    \label{fig:DissipVsRe}
\end{figure}

In the present case, the energy dissipation comes from two sources $\varepsilon=\varepsilon_a+\varepsilon_v$, where $\varepsilon_a$ denotes the dissipation coming from the
large scale shear, while $\varepsilon_v$ corresponds to the dissipation due to the vortons. Normalizing by $\sigma_a^3/L_s$, we then get:
\begin{equation}
    \dfrac{\left \langle \epsilon \right \rangle }{\sigma_a^3/L_s}= \dfrac{2\pi^2 L^3}{\mathrm{Re}} + \dfrac{1}{\mathrm{Re}}\left(\dfrac{L_s}{\sigma_a}\right)^2(3-\delta)\dfrac{5}{32} \left \langle
    \dfrac{\Gamma}{\eta^3}\right \rangle.
\end{equation}
The first term scales as the inverse of the Reynolds number and corresponds to the large-scale dissipation. The second term represents the contribution of the vortons. These two contributions are plotted in  Figure  \ref{fig:time_series} both in the laminar and the turbulent regime. Before the transition, the dissipation due to the large scale flow indeed dominates, while in the turbulent regime, the vortons dissipation  dominates. This two regimes pilot the behaviour of the total energy dissipation as a function of the Reynolds number, shown in Fig. \ref{fig:DissipVsRe}.

Before the transition, the normalized dissipation rate indeed behaves as the inverse of the Reynolds number, while in the turbulent regime, we observe a slower power law decay, with a higher dispersion, which may be due to slower convergence, see below. This second regime can be fitted with a power law with a small exponent $\alpha=-0.10$, that could be the signature of logarithmic corrections.

\begin{figure}
    \centering
    \includegraphics[scale=1.]{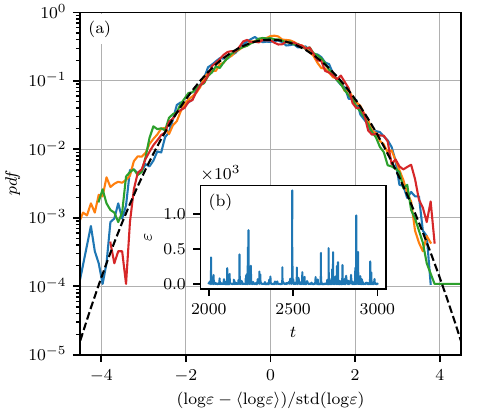}
    \caption{ (a) Standardized probability density function (pdf) of the dissipation rate logarithm for different Reynolds number in the turbulent state. The dashed black line correspond to to the normal distribution. The blue, orange, green and red lines respectively corresponds to $\mathrm{Re}=8.84\cdot 10^3; \, 9.84 \cdot 10^7; \, 1.32\cdot 10^{12}; \, 1.67\cdot10^{14}$. (b) Time series of the dissipation rate (for $\mathrm{Re}=1.32\cdot 10^{12}$), showing its highly intermittent dynamics characterized by strong bursts of dissipation separated by relatively quiescent time intervals.}
    \label{fig:pdf_diss}
\end{figure}

The scatter observed in the turbulent regime may be traced to the high intermittency of the energy dissipation, in analogy with what is observed in homogeneous isotropic  turbulence \cite{Frisch_1995}. In our case, again in agreement with homogeneous isotropic  turbulence\cite{Dubrulle_2019}, the statistics of energy dissipation
can be well approximated by a log-normal random distribution, see Fig. \ref{fig:pdf_diss}.

\section{Discussion}
We have introduced a new toy model of shear flows, exploiting the spatial intermittency and the scale separation between the large-scale flow and the small-scale structures. The model is very sparse, as only the most intense structures are considered, that are modeled via vortons, representing dynamically regularized quasi-singularities subject to rapid distorsion by the large-scale shear, and which retroact on this large-scale flow via the subgrid stress. The model displays an interesting transition between two regimes: (i) a laminar regime, in which all the dissipation is accounted for by the large-scale flow, and the vortons dynamics is essentially diffusive; (ii) a turbulent regime, in which most of the dissipation is produced by the vortons. These two regimes correspond to different scalings of the dissipation and the Grashof number as a function of Reynolds, with power laws that resembles the laws observed in classical turbulence. \

This shows that despite its simplicity, the new toy model may be of interest to understand or reproduce some of the observed properties of shear flows. As it stands, our model cannot be used directly for sub-grid modeling of shear flows, as it includes several arbitrary parameters that would need calibration against a DNS at least in the idealized case of  fully developed 3D turbulence with strong
imposed shear (e.g. flow over a flat plate, Couette flow). Using physical arguments, we tried to decrease as much as possible the number of free parameters in our model, but they remain actually two of them, namely the coupling parameter $\theta$ and the parameter controlling viscous dissipation $\delta$. In addition, there might exist an optimum for the number of vortons we need to use. We have found that increasing the density of vortons by a factor XX does not change the results, but there may exist an optimal value of the density to get better agreement with DNS. In addition, our representation of the feedback between vortons and the mean shear is limited to the case where the large scale flow is overdamped and does not change shape due to the interaction. A perhaps more realistic computation could include additional mode describing the large scale shear, at the expense of simplicity. Finally, we did not systematically vary the parameters of the forcing (its shape, and maximum frequency) to check its influence.\

In the spirit of understanding at least qualitatively what are the influence of various additional mechanisms on shear flow dynamics, one could however try to modify the toy model, to study specific effects. For example, in this preliminary validation, we neglected the feedback on the profile shape of the large-scale flow by introducing an ad-hoc forcing. This situation is perhaps more realistic in geophysical flows, where external forces such as solar irradiance and the Coriolis force determine velocity and temperature
gradient that do not deviate too much from quasi-geostrophy and adiabatic profile, at least in the midlatitudes. In the case of shear flows forced by boundary conditions
(Couette flow) or via a constant pressure gradient (Poiseuille), we know that this condition is  not realistic, as the turbulent fluctuations are known to flatten the global shear in the middle
of the domain. Even if we consider the layer just above the boundary layer, it is well known that the velocity profile switches from linear to logarithmic. As discussed in~\cite{Nazarenko_2000a,Nazarenko_2000,Dubrulle_2001}, this effect can be explained via rapid distorsion theory, that predicts that the $x-y$ component of the subgrid stress tensor scales inversely with the local large-scale shear, leading to the log-law of the wall after integration. Due to our approximation, we cannot capture this effect here, but it would be interesting to generalize our model to take into account the local shear. Note that all the computations made in~\cite{Nazarenko_2000a,Nazarenko_2000,Dubrulle_2001} use localized Gaussian wave-packets of vorticity, that are very close in spirit to our quasi-singularities. The main difference comes from our dynamical regularization which may introduce new effects.
Another interesting generalization would be towards geophysical flows, and especially localized extreme events such as convective storms. Indeed, individual convective cells are relatively sparse, and move within the "synoptic" (large-scale) wind and temperature fields, while interacting with nearest neighbors. If conditions are favorable, they can further organize into clusters known as mesoscale convective systems~\cite{Houze_2018}, which can produce significant hazard. An example of such severe storms are "derechos" which are long-lived MCS producing widespread severe surface wind gusts~\cite{Fery_2024}.
To deal with convective systems, one needs to add the coupling between temperature and velocity,
as well as moisture effect. Work is currently in progress to generalize our model to describe such type of coherent structures.


\section*{Acknowledgments}
This work received funding from the Ecole Normale Sup\'erieure de Lyon, from ANR TILT grant agreement no.
ANR-20-CE30-0035, from ANR BANG grant agreement no. ANR-22-CE30-0025, from the CNRS Program Recherche Risques ALEAS and from the CEA program Focus Numérique Frugal.

\appendix

\section{Kinetic energy}
\label{app:energy}
To derive the kinetic energy of the vortons field, we can first derive the expression for the energy spectrum of the vortons field,
\begin{equation}\label{eq:1DEnergySpec}
    E(\rho,t)=\dfrac{1}{2} \int_{\mathbb{S}^2(0, \rho)} \|\widehat{\veloc}(\bm{k},t)\|^2  \mathrm{d} \bm{k}.
\end{equation}
To do so, let us first write the Fourier transform of the vortons velocity field:

\begin{equation}\label{eq:veloc_fourier}
    \widehat{u}_\eta^m(\bm{k})=-\varepsilon_{m,n,p }  \dfrac{i}{4\pi} \dfrac{  k^n\eta K_1( \eta | \bm{k}|)}{|\bm{k}|}\sum_{\alpha=1}^N e^{-i \bm{k}\cdot \pos_\alpha} \gamma_\alpha^p,
\end{equation}

where $K_1$ is the second type  modified Bessel function of order one and $\varepsilon_{m,n,p}$ is the Levi-Civita symbol. We then obtain the one dimensional energy spectrum \eqref{eq:1DEnergySpec}:


\begin{widetext}
    \begin{equation}
        E(\rho,t)= \dfrac{ \left[\eta \rho  K_1(\eta \rho )\right]^2 }{2\pi}\left[ \dfrac{2}{3} \Gamma+  \sum_{\underset{\alpha\neq \beta}{\alpha , \beta}}  (\gams_\alpha\cdot\gams_\beta) \phi_1\left(2\pi \rho  \| \bm{r}_{\alpha \beta} \|\right)+\dfrac{\gams_\alpha\cdot \bm{r}_{\alpha \beta}}{\|\bm{r}_{\alpha \beta} \|}\dfrac{\gams_\beta \cdot \bm{r}_{\alpha \beta} }{\|\bm{r}_{\alpha \beta} \|} \phi_2\left(2\pi \rho  \| \bm{r}_{\alpha \beta} \|\right)  \right]
        \label{eq:EnerSpecVortons}
    \end{equation}
\end{widetext}

with $\Gamma=\sum_\alpha \| \bm{\gamma}_\alpha\|^2$ and

\begin{equation}
    \begin{cases}
        \phi_1(z) & =z^{-3}\left(  (z^2-1)\sin z +z \cos z\right),   \\
        \phi_2(z) & =z^{-3}\left( (3-z^2) \sin z -3z\cos z \right) . \\
    \end{cases}
\end{equation}
The total kinetic energy energy of the vortons induced field is given by the integral of Eq. \eqref{eq:EnerSpecVortons} along $\rho$. This integral cannot be computed analytically for a finite $\eta$ except for the first term, which is the dominant one for the total kinetic energy in the small $\eta$ limit since it is independant of $\rho$. The dominant contribution to kinetic energy thus writes
\begin{equation}
    K_v= \dfrac{\Gamma}{64\eta}.
\end{equation}

\section{Derivation of the subgrid stress tensor term in the large-scale flow amplitude equation}
\label{app:subgrid_stress}
As explained in section \ref{sec:shear_flow}, we obtain the time evolution of the large-scale shear amplitude by computing:
\begin{equation}
    k_s^{-1} (L^3/2)^{-1}\int_{\mathrm{V}} \eqref{eq:vorti_filtered} \cdot \mathbf{e}_y   \cos\left( k_s z\right) \;\mathrm{d}\pos.
    \label{eq:derivation_step}
\end{equation}
To compute specifically the contribution of the subgrid stress tensor, we begin by noting that
\begin{eqnarray}
    &&\int \left(\cos \left(k_s z \right) \mathbf{e}_y \right) \cdot \left( \nabla \times \left[ \nabla \cdot \tau_\ell \right] \right) \mathrm{d}x \nonumber\\
    &&=-k_s^2 \Re \left( \widehat{\tau_\ell}^{z,x}\left(k_s \mathbf{e}_z\right) \right),
    \label{eq:proj_tensor}
\end{eqnarray}
where we used two integration by parts, $\widehat{\tau_\ell}$ being the Fourier transform of the subgrid stress tensor whose components are written as exponents. As the only non-zero component of the large scale velocity field is $U^x$, we get
\begin{equation}
    \widehat{\tau_\ell}^{z,x} \approx \widehat{U^x_\ell u^z_{\eta,\ell}}-\widehat{(U^x u_\eta^z)_\ell} + \widehat{u^x_{\eta,\ell} u^z_{\eta,\ell}}-\widehat{(u_\eta^x u_\eta^z)_\ell}.
\end{equation}

With Eq. \eqref{eq:veloc_fourier} we have,
\begin{eqnarray}
    \widehat{U^x u^z_\eta}(\bm{k})&&= \dfrac{a}{2i}\biggl[\widehat{u}^z_\eta\left( \bm{k}-k_s\mathbf{e}_z\right) \nonumber\\
        && \quad -\widehat{u}^z_\eta\left( \bm{k}+k_s\mathbf{e}_z\right) \biggr]
\end{eqnarray}

which vanishes when evaluated at $\bm{k}\propto\mathbf{e}_z$ because $\varepsilon_{z,z,p}=0$ in \eqref{eq:veloc_fourier}. Assuming further that $\left(U u_\eta\right)_\ell \approx U_\ell u_{\eta,\ell}$, we are left with the computation of

\begin{eqnarray}
    \widehat{\tau_\ell}^{z,x}\left( k_s\mathbf{e}_z\right) && = \widehat{u_{\eta,\ell}^x u_{\eta,\ell}^z}\left(k_s\mathbf{e}_z\right) \nonumber\\
    &&\quad -\widehat{(u_{\eta}^x u_{\eta}^z)_\ell}\left(k_s\mathbf{e}_z\right).
    \label{eq:}
\end{eqnarray}
By definition of the filtered velocity field, writing $\bm{p}=k_s \mathbf{e}_z$, one has
\begin{equation}
    \widehat{\tau_\ell}^{z,x}\left( \bm{p}\right)= \int g_\ell(\bm{q},\bm{p})\widehat{u}^x_\eta(\bm{q})\widehat{u}_\eta^z(\bm{p}-\bm{q})\mathrm{d}\bm{q},
\end{equation}
where $g_\ell(\bm{q},\bm{p})=\left[\widehat{G}(\ell\bm{q})\widehat{G}(\ell(\bm{p}-\bm{q})) - \widehat{G}(\ell\bm{p})\right]$ and $G_\ell$ is the filtering function. Injecting the Fourier transform of the velocity field and rescaling the integration variable by $\eta$, one obtains,
\begin{eqnarray}
    &&\widehat{\tau_\ell}^{z,x}\left( \bm{p}\right)=\dfrac{\varepsilon_{1,m,n}\varepsilon_{3,\alpha,\beta}}{(4\pi)^2\eta}\sum_{j_1,j_2}  e^{-i\bm{p}\cdot \pos_{i_2}}\gamma_{j_1,n}\gamma_{j_2,\beta} \nonumber\\
    &&\times \int g_\ell\left(\dfrac{\bm{q}}{\eta},\bm{p}\right) \dfrac{q_m q_\alpha}{|\bm{q}||\eta\bm{p}-\bm{q}|}K_1\left( |\bm{q}|\right)K_1\left( |\eta\bm{p}-\bm{q}|\right) \nonumber\\
    && \times e^{-i\frac{\bm{q} \cdot \mathbf{r}_{j_1 j_2}}{\eta}} \mathrm{d}\bm{q}
\end{eqnarray}
where we have used $\varepsilon_{3,\alpha,\beta}\left(\eta p_\alpha -q_\alpha\right)=-\varepsilon_{3,\alpha,\beta} q^\alpha$ because $\bm{p}\propto \mathbf{e}_z$. Then, we choose to mollify at a scale $\ell=\eta$ such that $\ell \ll L_s$ and we keep only the leading order term in $\ell / L_s$. Moreover, considering that $\eta$ is small compared to the average distance between two vortons, we keep only the resonant term $j_1=j_2=j$ in the sum. This yields
\begin{eqnarray}
    \widehat{\tau_\ell}^{z,x}\left( \bm{p}\right)&&=\dfrac{\varepsilon_{1,m,n}\varepsilon_{3,\alpha,\beta}}{(4\pi)^2\eta}\sum_{j}  e^{-i \bm{p}\cdot \pos_{j}}\gamma_{j,n}\gamma_{j,\beta} \nonumber \\
    && \quad \times \int \left[\widehat{G}(\bm{q})^2 -1\right]\dfrac{q_m q_\alpha}{|\bm{q}|^2}K_1\left( |\bm{q}|\right)^2 \mathrm{d}\bm{q}.
\end{eqnarray}
Then, as $G$ is a real radial function, its Fourier transform is also isotropic, so we end up with
\begin{eqnarray}
    \widehat{\tau_\ell}^{z,x}\left( \bm{p}\right)&&=\dfrac{1}{4\pi\eta}\sum_{j}  e^{-i \bm{p}\cdot \pos_{j}}\gamma_{j,z}\gamma_{j,x} \nonumber\\
    && \quad \times \int_0^\infty \left[1 -\widehat{G}(r)^2 \right]r^2 K_1\left(r\right)^2 \mathrm{d}r.
    \label{eq:tau_fourier}
\end{eqnarray}
The value of the remaining integral depends on the choice of the mollifier, and is between $0$ and $3\pi^2/32$. We therefore introduce a parameter $\theta \in [0,1]$ which will play the role of a coupling parameter between small scales and large scales such that the integral is equal to $(3\pi^2/32) \theta$. Therefore,
\begin{equation}
    \Re\widehat{\tau_\ell}^{z,x}\left(k_s \mathbf{e}_3\right)=\dfrac{3\pi\theta}{128\eta}\sum_{\alpha=1}^N \gamma_{\alpha,z}\gamma_{\alpha,x} \cos\left( k_s z_\alpha\right),
    \label{eq:sub_grid}
\end{equation}
which yields the second term in the right-hand side of the amplitude equation \eqref{aequation} by substituting \eqref{eq:sub_grid} in \eqref{eq:proj_tensor} and dividing by $k_s (L^3/2)$.

\section{Expected dynamics in the laminar regime}
\label{app:laminar_regime}
We can compute analytically the expected dynamics of several variables related to the vortons in the laminar regime. Vortons are advected by the shear flow only \eqref{eq:LamiX} and in particular,
\begin{equation}
    z_\alpha(t)=z_\alpha(0).
\end{equation}

Writing $\psi(t)=\sqrt{1+ 4\nu \delta t/\eta_0^2}$, Eq.\eqref{eq:LamiEta} solves as

\begin{equation}
    \eta(t)=\eta_0 \psi(t),
\end{equation}
and for $i=1,2$,
\begin{equation}
    \gamma_{\alpha,i}(t)= \gamma_{\alpha,i}(0)\psi(t)^{\frac{3 (2 \delta - 5)}{2 \delta}},
\end{equation}

while
\begin{eqnarray}
    &&\gamma_{\alpha,z}(t) = \psi(t)^{\frac{3 (2 \delta - 5)}{2\delta}} \biggl( \gamma_{\alpha,z}(0) \nonumber\\
    &&  + \gamma_{\alpha,x}(0)\dfrac{2\pi}{L_s}\int_0^t a(s) \cos\left(\dfrac{2\pi}{L_s} z_\alpha(s) \right) \mathrm{d}s  \biggr).
\end{eqnarray}
Recalling that $z_\alpha(t)=z_\alpha(0)$ and taking random initial intensities following a uniform law given by $\gamma_{\alpha,i}(0) \hookrightarrow \mathcal{U}\left(-I/(2\sqrt{N}), I/(2\sqrt{N})\right)$, we end up with
\begin{eqnarray}
    \mathbb{E} \Gamma(t) &&= \dfrac{I^2}{4} \psi(t)^{\frac{3 (2 \delta - 5)}{\delta}} \nonumber\\
    && \quad \times \Biggl[1 + \frac{2}{2m+1}\dfrac{t^2}{\tau_\Gamma^2} \sum_{j=0}^{m} \mathrm{sinc}^2\left(\frac{j\pi t}{T_f}\right) \Biggr],
    \label{eq:expect_Gamma}
\end{eqnarray}
where $\mathrm{sinc}(x)=\sin x/x$ and $\tau_\Gamma= \sqrt{12}\pi L^3\nu/(L_s f_0)$ and we also assumed that $N\ge 3^3$. At large time, the sum in Eq. \eqref{eq:expect_Gamma} behaves approximately (graphically) as $1/2(1+T_f/t)$, while $\psi(t)\sim \sqrt{t}$.
We then get:
\begin{eqnarray}
    \mathbb{E} \Gamma(t)&& \underset{t\rightarrow \infty	}{\propto} t^{\frac{10\delta - 15}{2\delta}},\nonumber\\
    K_v(t) \sim   \mathbb{E} \Gamma(t)/\psi &&\underset{t\rightarrow \infty	}{\propto} t^{\frac{9\delta - 15}{2\delta}},\nonumber\\
    \Dot{K}_v\sim\Omega (t)\sim \mathbb{E} \Gamma(t)/\psi^3 &&\underset{t\rightarrow \infty	}{\propto} t^{\frac{7\delta - 15}{2\delta}}.
    \label{largetest}
\end{eqnarray}

\nocite{*}

\bibliography{apssamp}

@PREAMBLE{
 "\providecommand{\noopsort}[1]{}" 
 # "\providecommand{\singleletter}[1]{#1}%" 
}

@article{Novikov_1983,
  title   = {Generalized Dynamics of Three-Dimensional Vortex Singularities (vortons)},
  author  = {Novikov, E. A.},
  year    = {1983},
  journal = {Zhurnal Eksperimentalnoi i Teoreticheskoi Fiziki},
  volume  = {84},
  pages   = {975--981},
  issn    = {0044-4510}
}

@article{Winckelmans_1988,
  title   = {Weak Solutions of the Three-Dimensional Vorticity Equation with Vortex Singularities},
  author  = {Winckelmans, G. and Leonard, A.},
  year    = {1988},
  journal = {Physics of Fluids},
  volume  = {31},
  number  = {7},
  pages   = {1838},
  issn    = {00319171},
  doi     = {10.1063/1.866680}
}

@article{Winckelmans_1993,
  title   = {Contributions to {{Vortex Particle Methods}} for the {{Computation}} of {{Three-Dimensional Incompressible Unsteady Flows}}},
  author  = {Winckelmans, G.S. and Leonard, A.},
  year    = {1993},
  journal = {Journal of Computational Physics},
  volume  = {109},
  number  = {2},
  pages   = {247--273},
  issn    = {00219991},
  doi     = {10.1006/jcph.1993.1216}
}

@book{Cottet_2000a,
  title      = {Vortex Methods: Theory and Practice},
  shorttitle = {Vortex Methods},
  author     = {Cottet, G.-H. and Koumoutsakos, Petros D.},
  year       = {2000},
  publisher  = {Cambridge University Press},
  isbn       = {978-0-521-62186-1},
  lccn       = {QA925 .C68 2000}
}

@article{Onsager_1949,
  title     = {Statistical Hydrodynamics},
  author    = {Onsager, Lars},
  year      = {1949},
  journal   = {Il Nuovo Cimento (1943-1954)},
  volume    = {6},
  number    = {Suppl 2},
  pages     = {279--287},
  publisher = {Societ{\`a} Italiana di Fisica Bologna},
  issn      = {1827-6121}
}

@book{Frisch_1995,
  title      = {Turbulence: The Legacy of {{A}}.{{N}}. {{Kolmogorov}}},
  shorttitle = {Turbulence},
  author     = {Frisch, U.},
  year       = {1995},
  publisher  = {Cambridge University Press},
  isbn       = {978-0-521-45103-1 978-0-521-45713-2},
  lccn       = {QA913 .F74 1995}
}

@article{Dubrulle_2019,
  title   = {Beyond {{Kolmogorov}} Cascades},
  author  = {Dubrulle, B{\'e}reng{\`e}re},
  year    = {2019},
  journal = {Journal of Fluid Mechanics},
  volume  = {867},
  pages   = {P1},
  issn    = {0022-1120, 1469-7645},
  doi     = {10.1017/jfm.2019.98}
}

@article{Dubrulle_2001,
  title   = {A Dynamic Subfilter-Scale Model for Plane Parallel Flows},
  author  = {Dubrulle, B. and Laval, J. P. and Nazarenko, S. and Kevlahan, N. K.-R.},
  year    = {2001},
  journal = {Physics of Fluids},
  volume  = {13},
  number  = {7},
  pages   = {2045--2064},
  issn    = {1070-6631},
  doi     = {10.1063/1.1378038}
}

@article{Nazarenko_2000,
  title   = {Nonlinear {{RDT}} Theory of Near-Wall Turbulence},
  author  = {Nazarenko, S. and Kevlahan, N. K. -R. and Dubrulle, B.},
  year    = {2000},
  journal = {Physica D: Nonlinear Phenomena},
  volume  = {139},
  number  = {1},
  pages   = {158--176},
  issn    = {0167-2789},
  doi     = {10.1016/S0167-2789(99)00218-3}
}

@article{Nazarenko_2000a,
  title   = {Exact Solutions for Near-Wall Turbulence Theory},
  author  = {Nazarenko, S.},
  year    = {2000},
  journal = {Physics Letters A},
  volume  = {264},
  number  = {6},
  pages   = {444--448},
  issn    = {0375-9601},
  doi     = {10.1016/S0375-9601(99)00840-3}
}

@article{Sreenivasan_1984,
  title   = {On the Scaling of the Turbulence Energy Dissipation Rate},
  author  = {Sreenivasan, K. R.},
  year    = {1984},
  journal = {The Physics of Fluids},
  volume  = {27},
  number  = {5},
  pages   = {1048--1051},
  issn    = {0031-9171},
  doi     = {10.1063/1.864731}
}

@article{Eyink_2024,
  title   = {Onsager's `Ideal Turbulence' Theory},
  author  = {Eyink, Gregory},
  year    = {2024},
  journal = {Journal of Fluid Mechanics},
  volume  = {988},
  pages   = {P1},
  issn    = {0022-1120, 1469-7645},
  doi     = {10.1017/jfm.2024.415}
}

@article{Fery_2024,
  title      = {Analysing 23 Years of Warm-Season Derechos in {{France}}: A Climatology and Investigation of Synoptic and Environmental Changes},
  shorttitle = {Analysing 23 Years of Warm-Season Derechos in {{France}}},
  author     = {Fery, Lucas and Faranda, Davide},
  year       = {2024},
  journal    = {Weather and Climate Dynamics},
  volume     = {5},
  number     = {1},
  pages      = {439--461},
  publisher  = {Copernicus GmbH},
  doi        = {10.5194/wcd-5-439-2024}
}

@article{Houze_2018,
  title     = {100 {{Years}} of {{Research}} on {{Mesoscale Convective Systems}}},
  author    = {Houze, Robert A.},
  year      = {2018},
  journal   = {Meteorological Monographs},
  volume    = {59},
  number    = {1},
  pages     = {17.1-17.54},
  publisher = {American Meteorological Society},
  doi       = {10.1175/AMSMONOGRAPHS-D-18-0001.1},
  chapter   = {Meteorological Monographs}
}

@article{Mimeau_2021a,
  title = {A {{Review}} of {{Vortex Methods}} and {{Their Applications}}: {{From Creation}} to {{Recent Advances}}},
  author = {Mimeau, Chlo{\'e} and {Iraj Mortazavi} and Mortazavi, Iraj},
  year = {2021},
  journal = {Fluids},
  volume = {6},
  number = {2},
  pages = {68},
  doi = {10.3390/fluids6020068}
}

@article{Saffman_1986,
  title = {Difficulties with Three-Dimensional Weak Solutions for Inviscid Incompressible Flow},
  author = {Saffman, P. G. and Meiron, D. I.},
  year = {1986},
  journal = {Physics of Fluids},
  volume = {29},
  number = {8},
  pages = {2373},
  issn = {00319171},
  doi = {10.1063/1.865529}
}

@article{kornev2019large,
  title={Large eddy simulation with direct resolution of subgrid motion using a grid free vortex particle method},
  author={Kornev, N and Samarbakhsh, S},
  journal={International Journal of Heat and Fluid Flow},
  volume={75},
  pages={86--102},
  year={2019},
  publisher={Elsevier}
}

@article{mumford2012euler,
  title={On Euler's equation andEPDiff'},
  author={Mumford, David and Michor, Peter W},
  journal={arXiv preprint arXiv:1209.6576},
  year={2012}
}

@article{choquin1988analyse,
  title={Sur l'analyse d'une classe de m{\'e}thodes de vortex tridimensionnelles},
  author={Choquin, J-P and Cottet, G-H and Dautray, R},
  journal={Comptes rendus de l'Acad{\'e}mie des sciences. S{\'e}rie 1, Math{\'e}matique},
  volume={306},
  number={17},
  pages={739--742},
  year={1988}
}

@article{alvarez2024stable,
  title={Stable vortex particle method formulation for meshless large-eddy simulation},
  author={Alvarez, Eduardo J and Ning, Andrew},
  journal={AIAA Journal},
  volume={62},
  number={2},
  pages={637--656},
  year={2024},
  publisher={American Institute of Aeronautics and Astronautics}
}

\end{document}